\documentclass[11pt]{article}

\newcommand{\beq}{\begin{equation}} 
\newcommand{\eeq}{\end{equation}}
\newcommand{\beqs}{\begin{eqnarray}} 
\newcommand{\eeqs}{\end{eqnarray}}
\begin{document}
\begin{titlepage}
\vskip 2.5cm
\begin{center}
{\LARGE \bf Exact Superpotentials for Theories}\\
\smallskip
{\LARGE \bf with Flavors via a Matrix Integral}\\
\vspace{2.71cm}
{\Large
Riccardo Argurio,
Vanicson L. Campos, \\ 
\smallskip
Gabriele Ferretti and 
Rainer Heise}
\vskip 0.7cm
{\large \it Institute for Theoretical Physics - G\"oteborg University and \\
\smallskip
Chalmers University of Technology, 412 96 G\"oteborg, Sweden}
\vskip 0.3cm
\end{center}
\vspace{3.14cm}
\begin{abstract}
We extend and test the method of Dijkgraaf and Vafa
for computing the superpotential of ${\cal N} = 1$ theories
to include flavors in the fundamental representation of the gauge group.
This amounts to computing the contribution to the superpotential
from surfaces with one boundary in the matrix integral.
We compute exactly the effective superpotential
for the case of gauge group
$U(N_c)$, $N_f$ massive flavor chiral multiplets in the fundamental and one 
massive chiral multiplet in the adjoint, together with a Yukawa coupling.
We compare up to sixth-order with the result obtained by standard field theory 
techniques in the already non trivial case of $N_c=2$ and $N_f=1$.
The agreement is perfect.
\end{abstract}

\end{titlepage}

In a recent paper, Dijkgraaf and Vafa~\cite{DV} (building on previous
work \cite{previous})
have proposed a simple technique 
for computing the effective 
superpotential for the glueball
field $S = -{1\over 32 \pi^2}\mathrm{tr} W^\alpha W_\alpha$ in a large class of
${\cal N} = 1$ theories. For instance, 
in the case of a $U(N_c)$ gauge group and 
chiral fields $\Phi_i$ in the adjoint representation interacting with a
tree level superpotential $ W_{tree}(\Phi_i, \lambda_a)$, 
one is instructed to compute the
matrix integral to leading order in $N_c$:
\beq
    e^{-\frac{N_c^2}{S^2} {\cal F}_{\chi=2}(S, \lambda_a)} \approx
    \int\;d\Phi_i\; e^{-\frac{N_c}{S}  W_{tree}(\Phi_i, \lambda_a)},
\eeq
where we have denoted by $\lambda_a$ the coupling constants appearing in 
the superpotential.
The effective superpotential is, in this case \cite{DV}:
\beq
    W_{DV}(S, \Lambda, \lambda_a) = N_c S (-\log(S/\Lambda^3) + 1) +
       N_c \frac{\partial {\cal F}_{\chi=2}(S, \lambda_a)}{\partial S},
    \label{VYgen}
\eeq
where the presence on $N_c \frac{\partial}{\partial S}$ is justified by the
combinatorics of diagrams written on surfaces with spherical topology.
The first piece of the superpotential is the Veneziano-Yankielowicz
superpotential 
for pure $SU(N_c)$ SYM \cite{VY}, while the second piece
which starts with $O(S^2)$ terms gives the instantonic corrections.
In the case that the matrix model is integrable we can write the
exact effective superpotential in closed form, otherwise we can 
compute it at any given order in $S$. Recent checks and developments
of the conjecture have been performed in \cite{Chekhov}--\cite{Gorski}.

In the case of gauge groups $SO(N_c)$ or $Sp(N_c)$ there are also 
contributions from non orientable diagrams that can be written on 
the projective plane and their contribution $G_{\chi=1}(S, \lambda_a)$ will
appear in (\ref{VYgen}) without the factor 
$N_c \frac{\partial}{\partial S}$.

It is a natural step to extend the conjecture to theories including
matter, that is $N_f$ chiral multiplets in the fundamental.
This is implemented simply by including surfaces with boundaries.
To be specific, in the case of gauge group $U(N_c)$, for adjoint matter
$\Phi_i$ and fundamental matter $Q_f$ and $\tilde Q^f$ one should first
compute
\beq
    e^{-\frac{N_c^2}{S^2} {\cal F}_{\chi=2}(S, \lambda_a) -
       \frac{N_c}{S} {\cal F}_{\chi=1}(S, \lambda_a)} \approx
    \int\;d\Phi_i\; d Q_f\; d \tilde Q^f\; 
       e^{-\frac{N_c}{S}  W_{tree}(\Phi_i, Q_f, \tilde Q^f,\lambda_a)},
\eeq
and then write (the non-orientable contribution $G_{\chi=1}(S, \lambda_a)$
is absent in this case):
\beq
    W_{DV}(S, \Lambda, \lambda_a) = N_c S (-\log(S/\Lambda^3) +1) + 
       N_c \frac{\partial {\cal F}_{\chi=2}(S, \lambda_a)}{\partial S} +
        {\cal F}_{\chi=1}(S, \lambda_a).
\eeq

We are now going to test this conjecture. We take
a $U(N_c)$ gauge theory with one adjoint chiral multiplet $\Phi$ and $N_f$
chiral multiplets in the fundamental $Q_f$ and $\tilde Q^f$.
The tree level superpotential gives masses to all matter fields and
moreover there is a cubic coupling between the fundamentals
and the adjoint. All other possible couplings are turned off.
The tree level superpotential reads:
\beq
W_{tree}={1\over 2}M \mathrm{tr} \Phi^2 + m Q_f \tilde Q^f 
+ g Q_f\Phi \tilde Q^f ,
\eeq
where the flavor indices are summed while the color indices are not
written explicitly.

Since there are no self interactions of the adjoint field $\Phi$, all
diagrams with interactions 
will involve at least one flavor loop, that is a boundary.
The genus zero piece of the matrix integral reduces trivially
to the (one loop) vacuum amplitude of the adjoint field, which
enforces the matching of scales in the
Veneziano-Yankielowicz piece of the superpotential.
A similar factor is also present in the flavor integral with
analogous consequences. By working directly with the scale of the pure 
SYM theory we can concentrate on the interacting part which receives no 
contributions from genus zero.
To leading order in $N_c$, the matrix integral is thus saturated
by planar diagrams with one boundary, which sum up to ${\cal F}_{\chi=1}$.

The matrix integral can thus be easily performed.
Write:
\beq
Z=\int\;d\Phi\; d Q_f\; d \tilde Q^f\; e^{-{N_c\over S}\left(
{1\over 2}M \mathrm{tr} \Phi^2 + m Q_f \tilde Q^f 
+ g Q_f\Phi \tilde Q^f\right)}
= \langle e^{-{N_c\over S}g Q_f\Phi \tilde Q^f} \rangle,
\eeq 
where the correlators are normalized such that:
\beq
\langle Q_{\alpha f}\tilde Q^{\beta g}\rangle = {1\over m}{S\over N_c}
\delta^\beta_\alpha \delta^g_f, \qquad \qquad 
\langle \Phi^\alpha_\beta \Phi^\gamma_\lambda \rangle = {1\over M}{S\over N_c}
\delta^\alpha_\lambda \delta^\gamma_\beta.
\eeq
Expanding the exponential we have:
\beq
\langle e^{-{N_c\over S}g Q_f\Phi \tilde Q^f} \rangle = 
\sum_{k=0}^\infty {1\over (2 k)!} \left({g N_c \over S}\right)^{2k}
\langle (Q_f\Phi \tilde Q^f)_1 (Q_f\Phi \tilde Q^f)_2 \dots (Q_f\Phi \tilde 
Q^f)_{2k}\rangle, \label{expan}
\eeq
where we took into account that only correlators of an even number 
of fields $\Phi$ are non zero.

It is a simple combinatorial exercise to extract from (\ref{expan})
the coefficients of the connected planar diagrams with one boundary.
The different diagrams can be obtained first by contracting
the $Q$s and $\tilde Q$s in $(2k-1)!$ ways to give a single boundary,
and then connecting $2k$ points on the boundary through $k$ non intersecting
lines (the $\langle \Phi\Phi\rangle$ propagators).
The solution to this last combinatorial problem can be found in \cite{BIPZ},
Eq.~(31). The result for the free energy is:
\beq
\frac{N_c}{S} {\cal F}_{\chi=1}(S, g,m,M)=-N_f \sum_{k=1}^\infty
{(2k-1)!\over (k+1)! k!} \left({g N_c \over S}\right)^{2k}
\left({S\over m N_c}\right)^{2k} \left({S\over M N_c}\right)^k N_c^{k+1},
\eeq
which we can rewrite, for $\alpha={g^2\over m^2 M}$, as:
\beq
{\cal F}_{\chi=1}(S, \alpha)=-N_f \sum_{k=1}^\infty 
{(2k-1)!\over (k+1)! k!}
\alpha^k S^{k+1}.
\eeq
This expression can actually be summed to give:
\beq
{\cal F}_{\chi=1}(S,\alpha)=-N_f S \left[
{1\over 2} +{1\over 4\alpha S}(\sqrt{1-4\alpha S}-1) - \log\left(
{1\over 2}+{1\over 2}\sqrt{1-4\alpha S}\right) \right].
\eeq
We thus claim that
\beq
W_{DV}=N_c S (-\log(S/\Lambda^3) +1)+{\cal F}_{\chi=1}(S,\alpha)
\label{exact}
\eeq
is the exact superpotential for our theory with $N_f$ flavors
and the Yukawa coupling to the adjoint matter field. 

Our next task is to provide a purely field theory deduction
of the effective potential, and check that it matches the expression
computed through the matrix model. This task can be performed
by considering the system as an ${\cal N}=1$  deformation of an 
${\cal N}=2$ SYM with matter, and then deducing the low energy
superpotential through the Seiberg-Witten curve of the system.
While this is an interesting problem, it is already possible to prove the power
of the DV approach in the simplest case, that is $N_c=2$ and $N_f=1$.
This case already gives a non trivial result from the
matrix model, and we are going to show that it matches precisely
the field theory exact superpotential that we can simply
obtain by the standard techniques of \cite{ILS,Int}.
In turn, this is a strong support for the conjectured exact
superpotential (\ref{exact}).

Note that the superpotential
(\ref{exact}) does not seem to discern between $N_f<N_c$ and $N_f\geq N_c$,
while for instance the Seiberg-Witten curve of the related systems does.
A hint that our solution for the matrix integral could break down 
is that for $N_f\geq N_c\,$, additional boundaries start giving 
a large contribution to the integral, so that
the $\chi=1$ term is no longer singled out.

Let us write the first few terms of our exact superpotential
in the case $N_c=2$ and $N_f=1$:
\beqs
W_{DV}& =& 2S(-\log(S/\Lambda^3) + 1) \nonumber
 -{1\over 2}\alpha S^2
-{1\over 2}\alpha^2 S^3 -{5\over 6} \alpha^3 S^4 \\
& & -{7\over 4} \alpha^4 S^5 - {21\over 5} \alpha^5 S^6
-11 \alpha^6 S^7 + O(\alpha^7). \label{dvs}
\eeqs
Now we integrate $S$ out by setting $\partial_S W_{DV}=0$ and solve
for $S(\Lambda)$ as an expansion in $\alpha$.
Plugging back into (\ref{dvs}) we find:
\beqs
W_{DV}(\Lambda, \alpha)&=&2\Lambda^3\left[1 -{1\over 4}\alpha \Lambda^3
-{1\over 8}(\alpha \Lambda^3)^2 -{1\over 8}(\alpha \Lambda^3)^3 \right.
\label{dvl}\\
& & \left. -{21\over 128}(\alpha \Lambda^3)^4 -{1\over 4}(\alpha \Lambda^3)^5
-{429\over 1024}(\alpha \Lambda^3)^6 +O(\alpha^7)\right]. \nonumber
\eeqs

We can now set out to obtain the same effective superpotential
through an independent route. Consider the $U(2)$ theory with
an adjoint matter field $\Phi$ and $N_f=1$ chiral fields
in the fundamental $Q$ and $\tilde Q$. 
Moreover let us denote by $\tilde \Lambda$ the scale
of this theory (to be more precise, of the $SU(2)$ factor).
The tree level superpotential we introduce is:
\beq
W_{tree}={1\over 2}M \mathrm{tr} \Phi^2 + m Q \tilde Q 
+ g Q\Phi \tilde Q. \label{wtree}
\eeq
Let us first integrate out the adjoint field $\Phi$.
This trivially gives $\Phi=-{g\over M}\tilde Q Q$, and substituting
into (\ref{wtree}) we get, in the notation of \cite{Int}:
\beq
W_{tree,d}=mX - {1\over 2}{g^2\over M} X^2,
\eeq
where $X=Q \tilde Q$ is the gauge invariant meson field.

We now add the Affleck-Dine-Seiberg piece to the superpotential \cite{ADS},
taking into account the matching of the scales $\hat \Lambda^5=M^2
\tilde \Lambda^3$, where $\hat \Lambda$ is the scale of the
gauge theory with one flavor. The exact effective superpotential is thus
given by:
\beq
W_{eff}={\hat \Lambda^5 \over X} +mX - {1\over 2}{g^2\over M} X^2.
\label{weff}
\eeq
The absence of further corrections can be checked with the
fact that the above superpotential leads to the right Seiberg-Witten
curve, see for instance \cite{IS}.
We can now integrate out the massive meson $X$ and thus obtain
the effective superpotential for the low energy pure $SU(2)$ theory
(the $U(1)$ factor is now decoupled),
whose scale $\Lambda$ is matched to be $\Lambda^6=m\hat \Lambda^5$.
The scale $\Lambda$ is now the same as in (\ref{dvl}).

The condition for the extremum of $W_{eff}$ is:
\beq
ax^3-x^2+1=0,
\eeq
where $a\equiv{g^2\over m^2 M}\Lambda^3=\alpha \Lambda^3$ and
$x={m\over \Lambda^3}X$. The solution, expanded in powers of $a$, is:
\beq
x=1+{1\over 2}a +{5\over 8}a^2+ a^3+{231\over 128}a^4 +{7\over 2}a^5
+{7293\over 1024}a^6+O(a^7). \label{landala}
\eeq
Plugging (\ref{landala}) back into the effective superpotential
(\ref{weff}), we get:
\beq
W_{eff}=2\Lambda^3\left(1-{1\over 4}a -{1\over 8}a^2
-{1\over 8}a^3 -{21\over 128}a^4 -{1\over 4}a^5-{429\over 1024}a^6
+O(a^7) \right),
\eeq
which agrees exactly with (\ref{dvl}).

\section*{Acknowledgments}
We would like to thank G.~Bonelli and P.~Salomonson
for discussions. 
This work is partly supported by EU contract HPRN-CT-2000-00122.

\end{document}